# Nonlinear optical effects of ultrahigh-*Q* silicon photonic nanocavities immersed in superfluid helium

Xiankai Sun, Xufeng Zhang, Carsten Schuck, and Hong X. Tang*

*Department of Electrical Engineering, Yale University, 15 Prospect Street, New Haven, Connecticut 06511, USA*

*hong.tang@yale.edu

**Photonic nanocavities are a key component in many applications because of their capability of trapping and storing photons and enhancing interactions of light with various functional materials and structures. The maximal number of photons that can be stored in silicon photonic cavities is limited by the free-carrier and thermo-optic effects at room temperature. To reduce such effects, we performed the first experimental study of optical nonlinearities in ultrahigh-*Q* silicon disk nanocavities at cryogenic temperatures in a superfluid helium environment. At elevated input power, the cavity transmission spectra exhibit distinct blue-shifted bistability behavior when temperature crosses the liquid helium lambda point. At even lower temperatures, the spectra restore to symmetric Lorentzian shapes. Under this condition, we obtain a large stored intracavity photon number of about 40,000, which is limited ultimately by the local helium phase transition. These new discoveries are explained by theoretical calculations and numerical simulations.**

## Introduction

Optical micro- and nanocavities are an essential building block in many applications that exploit light–matter interaction. Pedestal-supported microdisk cavities have been widely employed in recent studies of cavity optomechanics because of their ultrahigh optical quality factor (*Q*) and large coupling between optical and mechanical degrees of freedom [1-4]. By reducing the device size or the modal volume (*V*), one obtains the mechanical vibration modes at higher frequencies that could more easily enter the quantum regime with the optomechanical interaction [5]. The ultrahigh optical *Q* and small modal volume result in high intracavity energies and thus strong light–matter interaction. However, this also leads to a reduced threshold for undesired optical nonlinearities in particular the thermal effects that also depend inversely on $Q/V$. In silicon, these nonlinearities become appreciable at elevated optical power levels, as the Kerr effect changes the refractive index and two-photon absorption (TPA) generates free carriers, causing subsequent free-carrier dispersion (FCD) and free-carrier absorption (FCA). The heat generated from TPA and FCA results in temperature rise and additional refractive index change by the thermo-optic effect. All these nonlinear effects combine and can lead to bistability or self-pulsation phenomena in high-*Q* silicon cavities [6-12]. In recent race toward ground-state cooling of nanomechanical resonators, these undesired optical nonlinearities have also imposed a practical limit on the number of photons inside a cavity to achieve the strong optomechanical coupling [5,13,14].

Here, we investigate the nonlinear effects in ultrahigh-*Q* silicon disk cavities immersed in liquid helium (LHe). Cooling the devices to the helium lambda point (~ 2.17 K at 1 atm) and below ensures an ultrastable ambient temperature, as the surrounding helium transitions from a normal-fluid to a superfluid phase, resulting in an extremely high thermal conductivity (>$10^5$ W $m^{-1}$ $K^{-1}$, Ref. 15) and large specific heat capacity. In addition to a greatly reduced thermal resistance of the disk cavities, the thermo-optic coefficient of silicon also drops by at least 4 orders of magnitude from that at room temperature [16]. The combination of an ultrastable ambient temperature and the significantly reduced thermo-optic coefficient contributes to the vanishing of the thermo-optic effect. On the other hand, as compared to room-temperature operation the TPA-induced free-carrier generation in silicon is largely suppressed at liquid helium temperatures because of the significantly reduced free-carrier lifetime [17]. In this paper, we present spectroscopy measurement results on optical transmission of our disk cavities. When immersed in superfluid helium, these devices demonstrate strikingly new behaviors compared with their normal operation in air: the transmission spectra exhibit blue-shifted bistability as the disk cavities are at 2.168 K near the liquid helium lambda point and then transition to more symmetric Lorentzian shapes at even lower temperatures. These features are well reproduced in numerical calculations by introducing the relevant nonlinear effects into the coupled-mode theory. In the superfluid regime below the lambda point, the cavity field could build up to sustain about 40,000 photons without the worry of the instability effects. Achieving a remarkably large cavity field otherwise unobtainable, this methodology paves the way for the study of cavity quantum optomechanics at cryogenic temperatures where high photon occupation, low phonon occupation, and low phonon dissipation are required.

## Results

As shown in Fig. 1, at the heart of our devices is a 1.16-μm-radius disk cavity surrounded by a suspended sunflower-type circular photonic crystal, fabricated on an all-integrated silicon-on-insulator substrate with CMOS-compatible processes. These cavities exhibit measured intrinsic (radiation) quality factors from $3\times10^5$ to $5\times10^5$ with a cavity modal volume as small as $2.1(\lambda_0/n_0)^3$, where $\lambda_0$ is the cavity resonant wavelength and $n_0$ the refractive index of silicon. Light is coupled into and out of the cavity with photonic crystal access waveguides connected via strip waveguides to a pair of grating couplers for out-of-plane optical access. The chip is mounted inside the sample chamber of a helium-4 cryostat, which is equipped with needle valves for controlling the liquid helium content inside the sample chamber. Therefore the devices can be measured in vacuum as well as in normal-fluid or superfluidic helium.

We perform power-dependent transmission spectroscopy measurements on the devices. Light from a tunable laser is polarization-adjusted and then delivered to the device via an optical fiber array. The optical power coupled into the feeding waveguide is controlled with an optical attenuator. The output light is collected by a low-noise photodetector to record the transmission spectra. Note that the as-measured transmission includes the insertion loss from the two grating couplers over the measured wavelength range (21–27 dB). We characterize two different device designs: Device 1 shows low transmission as the disk cavity and the photonic crystal access waveguide are undercoupled, while Device 2 has higher transmission as a result of slight overcoupling between the cavity and waveguide.

Due to the small device size, the disk cavities exhibit a single resonance peak over a wavelength range of tens of nanometers [1,18]. Figure 2 shows the zoomed optical transmission spectra of an undercoupled device (Device 1) measured under three different conditions: (a) in air at room temperature (298 K), (b) immersed in superfluid helium near the lambda point (at 2.168 K) and (c) far below the lambda point (at 2.000 K). It should be noted that despite the different environmental conditions the modal distribution of the photonic structures remains essentially the same in liquid helium as in air or vacuum because of its gas-like refractive index (1.026).

At low optical input power (below −20 dBm) the cavity is still in linear regime, and we obtain the disk's resonant wavelength ($\lambda_0$) and loaded quality factor ($Q_L$) by a Lorentzian fit to the transmission spectra for each of the three measurement condition. Compared to the results obtained in air, the cavity resonance shifts toward the short wavelength side by 11.4 nm at liquid helium temperatures because of the reduced refractive index of silicon [19]. The coupling quality factor ($Q_{in}$) due to the in-plane loss to the access waveguides can be inferred from the normalized cavity transmission $T = (Q_L/Q_{in})^2$. The intrinsic quality factor ($Q_v$) due to vertical radiation loss is then determined by $Q_v = Q_{in}Q_L/(Q_{in} - Q_L)$. These measured cavity properties are listed in Table I, where we find an enhancement of the intrinsic quality factor when the disk is immersed in liquid helium, which we attribute to reduced surface scattering loss.

At high optical input power the cavity enters the nonlinear regime. Here the transmission spectra show different behavior under the three measurement conditions given above: At room temperature in air we observe a typical red-shifted bistability [Fig. 2(a)], while the spectra taken in superfluid helium near the lambda-point temperature exhibit blue-shifted bistability [Fig. 2(b)]. However, when the disk cavity is immersed in superfluid helium at an even lower temperature 2.000 K, the spectra remain almost symmetric with Lorentzian shapes [Fig. 2(c)].

Figure 2(d) shows the observed shift of the peak wavelength as a function of the input power for the three cases.

The nonlinear behavior of the cavity can be explained by the influence of the air/helium environment on the free-carrier and thermo-optic effects. Due to the centrosymmetry of crystalline silicon, the second-order nonlinearities are inhibited. Therefore, the third-order nonlinearities dominate at elevated power levels, resulting in cavity dispersion (by FCD, thermo-optic effect, and Kerr effect) and additional optical loss (by TPA and FCA). Taking all these effects into account, we find the steady-state characteristic equation by applying the coupled-mode theory to a two-port system consisting of a cavity and a coupling bus waveguide [9,20]:

$$\begin{aligned}\frac{\omega_0'}{2Q_{in}}|s_1|^2 &= |a|^2(\omega-\omega_0')^2 \\ &+\frac{|a|^2}{4}\left[\frac{\omega_0'}{Q_v}+\frac{\omega_0'}{Q_{in}}+\frac{\beta c^2}{n^2V_{TPA}}|a|^2+\frac{e^3}{n^2\omega^2\varepsilon_0}\left(\frac{\beta c^2}{n^2V_{TPA}}\frac{|a|^4}{2\hbar\omega}\frac{\tau_{recon}}{V_{FC}}\right)\left(\frac{1}{m_e^{*2}\mu_e}+\frac{1}{m_h^{*2}\mu_h}\right)\right]^2,\end{aligned} \tag{1}$$

where the cavity resonant wavelength including nonlinear effects is

$$\begin{aligned}\frac{2\pi c}{\omega_0'} &= \lambda_0-\frac{\lambda_0 e^2}{2n_0 n\omega^2\varepsilon_0}\left(\frac{\beta c^2}{n^2V_{TPA}}\frac{|a|^4}{2\hbar\omega}\frac{\tau_{recon}}{V_{FC}}\right)\left(\frac{1}{m_e^*}+\frac{1}{m_h^*}\right)+\frac{\lambda_0}{n_0}\frac{\partial n}{\partial T}|a|^2 R \\ &\times\left[\frac{\beta c^2}{n^2V_{TPA}}|a|^2+\frac{e^3}{n^2\omega^2\varepsilon_0}\left(\frac{\beta c^2}{n^2V_{TPA}}\frac{|a|^4}{2\hbar\omega}\frac{\tau_{recon}}{V_{FC}}\right)\left(\frac{1}{m_e^{*2}\mu_e}+\frac{1}{m_h^{*2}\mu_h}\right)\right]+\frac{\lambda_0}{n_0}\frac{n_2 c}{nV_{Kerr}}|a|^2.\end{aligned} \tag{2}$$

In Eqs. (1) and (2), the cavity field amplitude $a$ is normalized such that its squared magnitude $|a|^2$ equals the cavity energy. Likewise, $s_1$ is the waveguide field amplitude with $|s_1|^2$ being the input power in the bus waveguide. The constants $c$, $e$, $\varepsilon_0$, and $\hbar$ are the speed of light in vacuum, the elementary charge, the vacuum permittivity, and the reduced Planck constant, respectively. All other material and cavity parameters are listed in Table 1.

The group index $n$ is taken to be equal to the material index $n_0$ by neglecting the waveguiding and material dispersion effects [7]. The cavity volume for TPA is calculated for our disk devices by a finite-difference time-domain (FDTD) method [7,9,10]. The cavity volume for the Kerr effect $V_{\mathrm{Kerr}}$ is set to be that of TPA, $V_{\mathrm{TPA}}$. The cavity volume for free carriers $V_{\mathrm{FC}}$ is also calculated based on a definition from Ref. 7 which neglects the nonlocal effects due to spatial carrier diffusion. The carrier mobilities are generally concentration and temperature dependent but converge to constant values at very high carrier concentrations [21-24]. Therefore we use their convergent values in our model for simplicity. The carrier recombination time is *a-priori* unknown due to its dependence on the structural geometry, temperature, and carrier concentration. Here we neglect its carrier concentration dependence and obtain its values under different measurement conditions from the optimal fit to our experimental curves taking into account its temperature dependence [17]. The disk's thermal resistance $R$ is treated as a second fit parameter which characterizes the temperature change due to the TPA- and FCA-generated heat. Its value depends on the cavity geometry, the thermal conductivity and heat capacity of both the cavity and environment. When measured in air, the disk cavity is more thermally isolated from its surrounding photonic crystal membrane and the underlying substrate. The optimal fit yields a large $R$ value of 530 K $\mathrm{mW}^{-1}$, which is similar to that of a ladder-shaped photonic crystal beam cavity [11]. As the disk is immersed in superfluid helium, however, the $R$ value cannot be experimentally extracted anymore, since the thermo-optic effect is highly suppressed due to the vanishing thermo-optic coefficient $\partial n/\partial T$ at this temperature. In this regard, the third term on the right-hand side of Eq. (2) does not play a role, and we may as well use a typical value for bulk silicon (20 K $mW^{-1}$) for the calculation.

With these parameters, the spectra of the cavity energy $|a|^2$ as calculated from Eqs. (1) and (2) are displayed in Fig. 3(a)–(c) for different power levels. We find good agreement with the experimental data (see Fig. 2). Since the temperature change is not reflected in the transmission spectra at liquid helium temperatures due to the vanishing coefficient $\partial n/\partial T$, we performed finite-element simulations to obtain the temperature profiles and compare them in air and in superfluid helium. Figure 3(d) shows cross-sectional profiles of the temperature distribution $\Delta T$ in units of kelvin for the same optical mode measured in the experiments. The surrounding photonic crystal structures are omitted in the plots for clarity. It is evident that with the same amount of circulating optical power in the disk cavity, the temperature rise for the disk placed in air is about 1700 times higher than that in a superfluid helium environment. A second feature is a much more localized temperature gradient inside the disk when it is placed in liquid helium. We attribute this to the high thermal conductivity of the superfluid helium which helps to dissipate the generated heat very efficiently at the surfaces of the disk cavity.

Our model thus explains how the free-carrier and thermo-optic effects are affected by the liquid helium environment: The red-shifted bistability observed in air at room temperature is due to the dominating thermo-optic effect on the thermally isolated disk cavity. When the disk is immersed in superfluid helium at 2.168 K, the efficient heat dissipation at the cavity surfaces and the vanishing thermo-optic coefficient together contribute to the disappearance of the thermo-optic effect, and thus the remaining FCD is responsible for the blue-shifted bistability. As the device is cooled further down to 2.000 K, the carrier lifetime is further reduced such that the FCD effect diminishes, and thus more symmetric Lorentzian shapes of the cavity resonance are maintained even at elevated optical power levels. From the experiment and calculation, we estimate a maximum of about 40,000 photons inside the disk cavity without the presence of the undesired bistability effects. This intracavity photon number presents at least an order of magnitude enhancement over that achieved in Ref. 5. The reduction of carrier lifetime from 2.168 K to 2.000 K should be attributed to a better conductivity of the ambient superfluid helium, as the ion mobilities start to grow nearly exponentially with temperature dropping from the lambda point [25].

In the above experiments performed at 2.000 K in superfluid helium, the highest input power is limited to 3 dBm. For an undercoupled cavity device (Device 1), the majority of the optical power is not delivered to the disk cavity, but converts into heat inside the sample chamber due to scattering at the grating couplers and photonic crystal waveguides. The heat generation becomes appreciable once the heat load exceeds the cooling capacity of the cryostat. Such effects manifest as an overall blue-shift of the cavity resonance albeit preserving near-symmetric Lorentzian shapes as shown in Fig. 2(c) for input power of 0 and 3 dBm. Minimizing the perturbation to the environment, we measure an overcoupled disk cavity (Device 2) to study the cavity response in liquid helium at 2.000 K. Its transmission spectra for input power up to 6 dBm are displayed in Fig. 4(b), where the quality factors $Q_v = 3.02\times10^5$ and $Q_{in} = 1.59\times10^5$ are extracted from low-power measurements. The spectra from the same device collected in vacuum at 77.7 K are provided in Fig. 4(a) for comparison. The spectra in Fig. 4(b) show a slight blue-shift of the resonance peaks with increasing input power until they are eventually clamped at ~2.5 μW of transmitted power for input power levels of more than 0 dBm. Such transmission clamping is also observed from many other devices and is not a result of saturation of our photodetectors. We attribute this effect to the self-regulation of the power inside the cavity: during the spectrum measurement, when the input laser wavelength approaches the cavity resonance, the intracavity field builds up and the absorption-induced heat is dissipated to the ambient, causing a local phase transition of superfluid helium to its normal state. In the normal-fluid helium, the device becomes susceptible to the stronger free-carrier effects again, which in turn reduces the intracavity power. We have experimentally verified that the cavity transmission in normal-fluid helium is indeed much lower than that in superfluid helium (see the results from a strongly overcoupled device in Supplementary Information).

## Discussion

We have conducted the first spectroscopic studies of the optical nonlinearities of wavelength-sized ultrahigh-$Q$ silicon disk cavities immersed in superfluid helium, where the free-carrier and thermo-optic effects are greatly suppressed. In stark contrast to the red-shifted bistability usually observed for devices in air or vacuum at room temperature, the cavity transmission spectra exhibit blue-shifted bistability when devices are immersed in superfluid helium at 2.168 K (near the liquid helium lambda point) and then transition to more symmetric Lorentzian shapes at a lower temperature 2.000 K. The cavity transmission keeps increasing with the increased input power without significant $Q$ degradation until it is fully clamped due to a self-regulation mechanism induced by the local phase transition of the ambient liquid helium. With a nonlinear coupled-mode theory model, we derived the steady-state characteristic equation for calculating the transmission spectra. The major features of the experimental spectra are well reproduced in the calculations and the maximal number of photons inside the cavity is found to be around 40,000. By demonstrating the interesting nonlinear phenomena from silicon disk cavities, our work opens the door for the study of integrated nonlinear photonics in a liquid helium environment. This methodology can directly be applied to cavity quantum optomechanics by taking advantage of the high thermal conductivity, large specific heat capacity, low density, and low viscosity of liquid helium. Further cooling the superfluid deep into sub-kelvin regime will completely eliminate the dynamic viscosity of superfluid [26]. This will bring cavity optomechanics to a brand new regime where high photon occupation and low mechanical damping can simultaneously be realized.

## Methods

### Sample design and fabrication

The disk nanocavities in our experiment possess simultaneous ultrahigh $Q$ factor and small modal volume. This is accomplished by embedding the wavelength-sized disk inside a sunflower-type circular photonic crystal, the bandgap of which structure provides excellent confinement to the disk's whispering-gallery modes and leads to their ultrahigh $Q$ factors beyond that of bending-loss limit. Structure optimization is performed with MEEP by a three-dimensional FDTD method [27], where the disk radius, the photonic crystal's lattice constant and hole radius, and the gap between the disk and the photonic crystal are all optimized in a multi-variable parameter space to maximize the cavity $Q$ factor [18].

The optimized geometry is fabricated on a standard silicon-on-insulator substrate with 220-nm silicon layer on 3-μm buried oxide. The pattern is first defined by electron-beam lithography, and then transferred to the silicon layer through chlorine-based inductively coupled plasma reactive ion etching. With a photolithography step and subsequent wet etching, the underlying buried oxide is removed such that the circular photonic crystal is completely released from the substrate while an oxide supporting pedestal still remains beneath the disk.

**Immersing sample in superfluid helium**

We pump on the sample chamber and use a needle valve to control the flow of liquid helium from the dewar into the sample chamber. Several RF sensors are mounted inside the chamber to monitor the liquid helium level and ensure immersion of the chip. The vacuum pump effectively provides the major cooling power for the liquid helium inside the sample chamber and maintains its temperature lower than the outside helium bath. Further cooling and fine temperature control is provided by pumping on an additional 1K-pot (not shown in Fig. 1). By balancing the pumping power and helium flow into both the sample chamber and the 1K-pot we are able to adjust the sample chamber temperature down to 2 K with a precision better than 1 mK.

**Simulation of the disk's temperature profile**

The temperature profiles of the disk cavity are simulated with COMSOL Multiphysics by a three-dimensional finite-element method. We first obtain the optical mode distribution of the disk cavity. The optical mode produces resistive heating that is proportional to the optical intensity. This spatial heat distribution is then used as the initial condition to obtain the temperature profile at thermal equilibrium. We compare two situations where the disk is placed in air at room temperature or immersed in superfluid helium at 2 K, assuming the same electrical conductivity $\sigma$ of 0.6 S $m^{-1}$ for silicon in both cases. We then obtain the temperature profiles in Fig. 3(d) for an intracavity optical energy of 1 fJ. The heat dissipation and temperature profiles are mainly determined by the thermal conductivity ($K$) and heat capacity ($C_p$), where the following values are used in our simulation: air [$K$ = 0.0257 W m$^{-1}$ K$^{-1}$, $C_p$ = 1005 J kg$^{-1}$ K$^{-1}$], liquid helium [$K \sim 10^5$ W m$^{-1}$ K$^{-1}$, $C_p$ = 5233 J kg$^{-1}$ K$^{-1}$], and silicon [at 298 K, $K$ = 148 W m$^{-1}$ K$^{-1}$, $C_p$ = 715 J kg$^{-1}$ K$^{-1}$; at 2 K, $K$ = 45 W m$^{-1}$ K$^{-1}$, $C_p$ = 2.21×10$^{-3}$ J kg$^{-1}$ K$^{-1}$, Ref. 28].

## Acknowledgements

This work was supported by Defense Advanced Research Projects Agency (DARPA) managed by Dr. J. R. Abo-Shaeer under the ORCHID program (contract number C11L10831), a STIR program managed by Dr. Samuel Stanton from Army Research Office (contract number J00212), and National Science Foundation CAREER award. Facilities used were supported by Yale Institute for Nanoscience and Quantum Engineering and NSF MRSEC DMR 1119826. H.X.T. acknowledges support from a Packard Fellowship in Science and Engineering. The authors thank Michael Power and Dr. Michael Rooks for assistance in device fabrication.

## Author contributions

X.S. performed the measurements, theoretical modeling, and numerical analysis under the supervision of H.X.T.; X.S. and X.Z. contributed to device design and fabrication; X.Z. performed the FDTD and finite-element simulation of the disk cavity's optical parameters and temperature profiles; C.S. assisted with the operation of cryostat during the cryogenic measurement; X.S. wrote the manuscript with the assistance of all the coauthors.

## Additional information

**Supplementary information** accompanies this paper at [URL provided by the publisher].

**Competing financial interests:** The authors declare no competing financial interests.

## Figures:

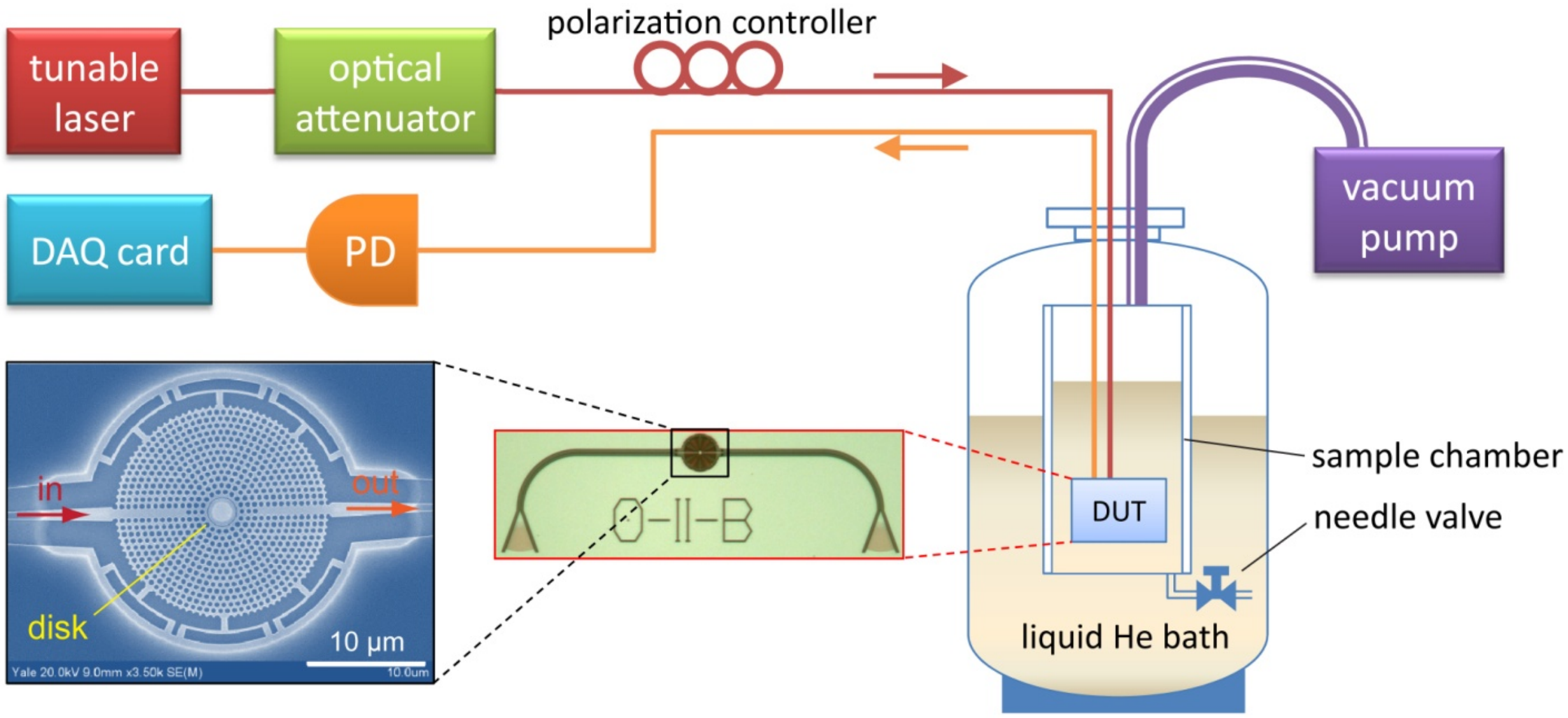

Figure 1. Schematic of the cryogenic measurement setup. The device under test (DUT) is placed inside the sample chamber in a liquid helium (He) cryostat. Light is coupled into and out of the device via an optical fiber array. The sample chamber is connected at the top to a vacuum pump and at the bottom to the liquid He bath with a control needle valve. With a careful balance of pumping power and liquid He flow rate, the liquid He inside the sample chamber can be maintained at a temperature below that of the He bath (4.2 K). An optical and SEM image of an example device is shown in the lower left panel. It consists of a pair of in- and output grating couplers, a wavelength-sized disk cavity embedded in a suspended circular photonic crystal, connecting strip waveguides, and photonic crystal access waveguides. DAQ: data acquisition. PD: photodetector.

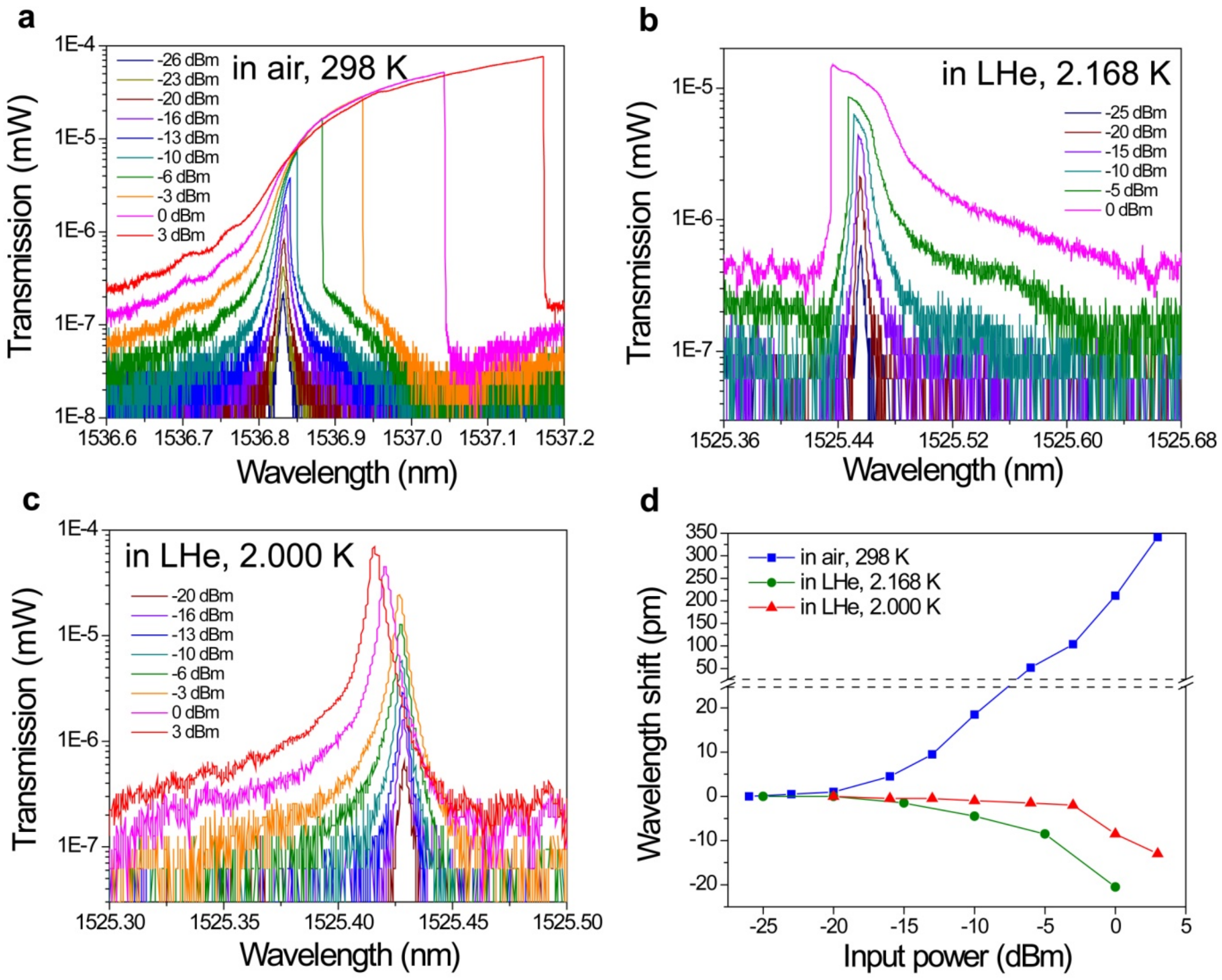

Figure 2. (a)–(c) Measured spectra of optical transmission through the disk cavity (Device 1) at various input power levels: (a) in air at 298 K (room temperature); (b) immersed in superfluid helium at 2.168 K; (c) immersed in superfluid helium at 2.000 K. The quoted input power values are measured before the input grating coupler. The transmitted power values are measured after the output grating coupler. The insertion loss induced by the pair of grating couplers is 21 dB in the measurements of (a) and (b), and 27 dB in (c). (d) Power dependence of the transmission peak wavelength for the three measurement conditions.

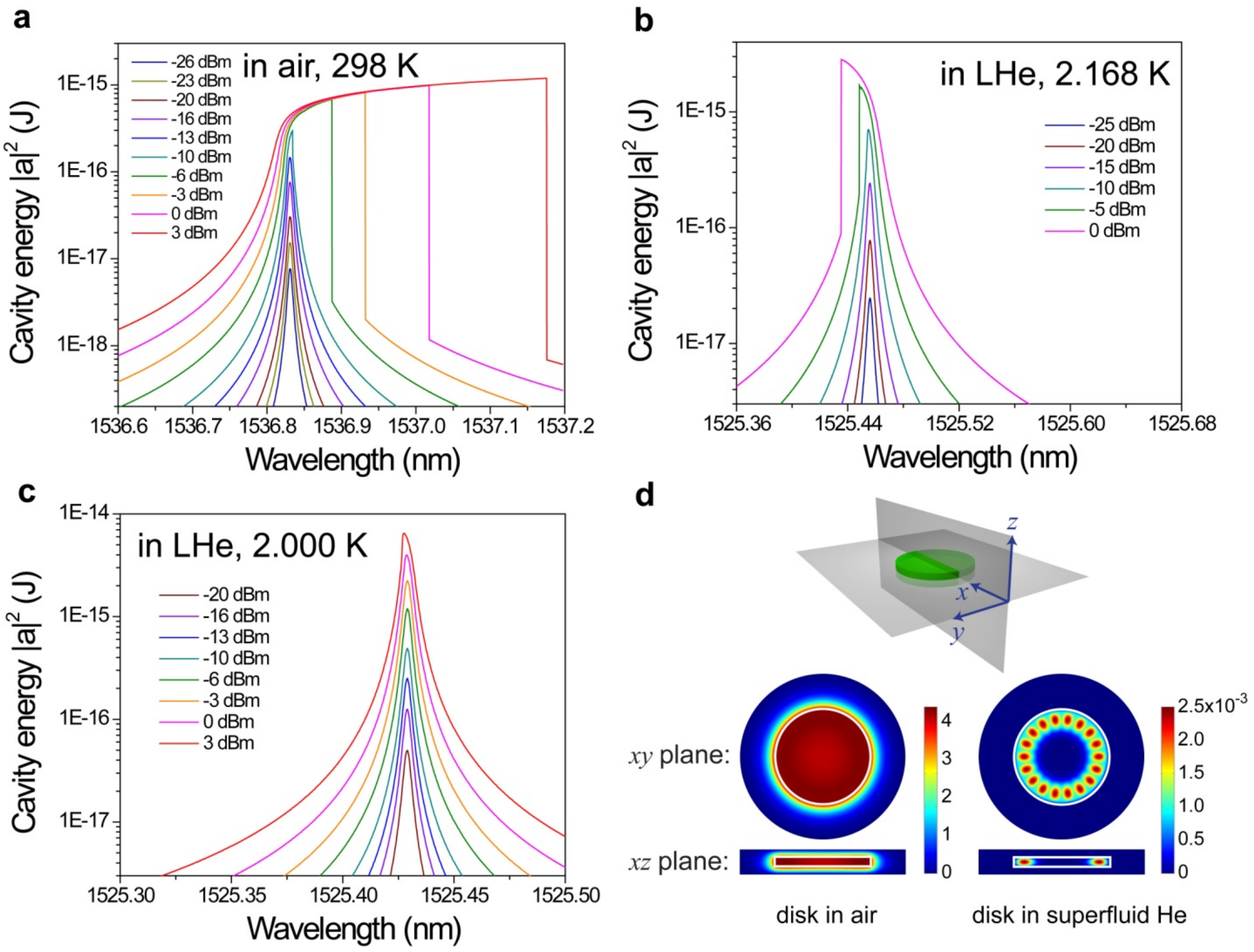

Figure 3. (a)–(c) Calculated energy spectra of the disk cavity (Device 1) at various input power levels: (a) in air at 298 K (room temperature); (b) in superfluid helium at 2.168 K; (c) in superfluid helium at 2.000 K. The quoted input power values correspond to those in the optical fiber before the input grating coupler in the experiments. (d) Comparison of the disk cavity temperature profiles ΔT (unit: K) from finite-element simulations for the same optical mode as measured in the experiments, in air at room temperature and in superfluid helium at 2 K, for the same amount of circulating optical power. The white boundaries mark the position of the disk cavity. The surrounding photonic crystal structures are omitted for clarity.

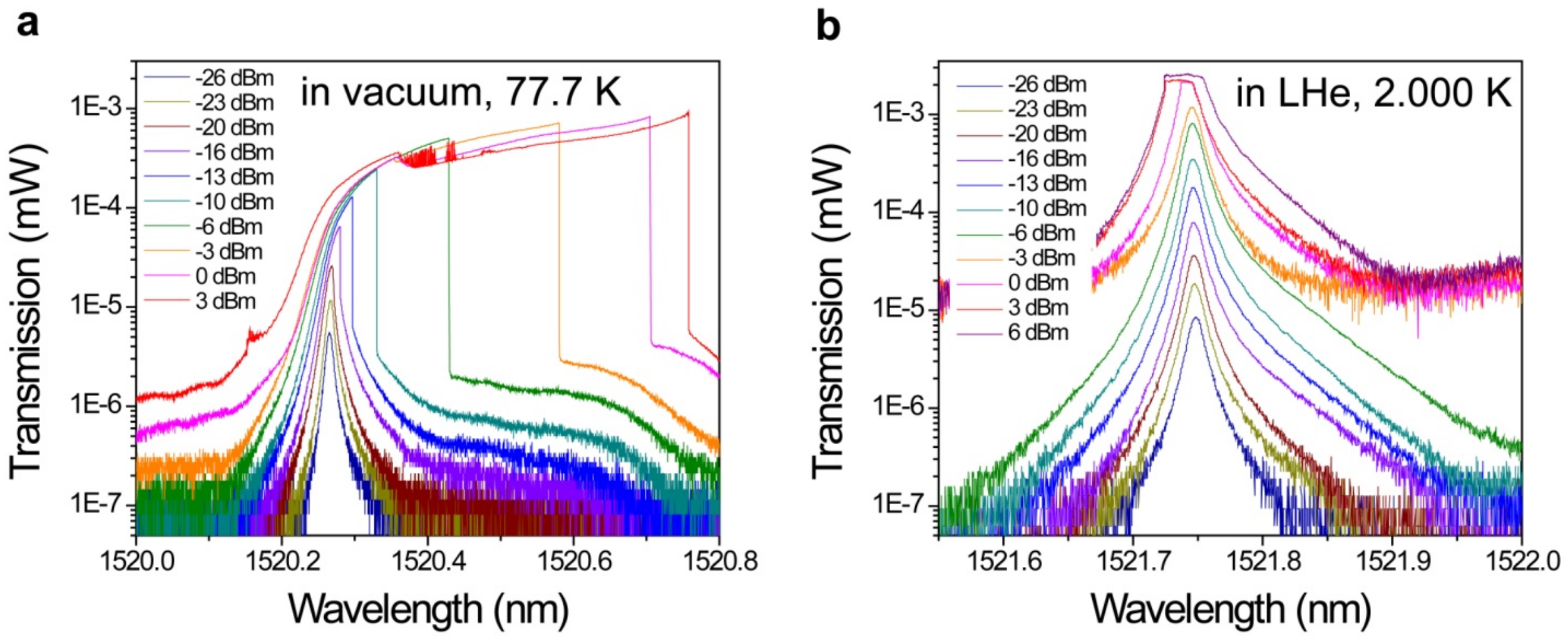

Figure 4. Measured spectra of optical transmission through the disk cavity (Device 2) at various input power levels: (a) in vacuum at 77.7 K; (b) immersed in superfluid helium at 2.000 K. The transmission is clamped at ~2.5 μW for input power higher than or equal to 0 dBm. The quoted input power values are measured before the input grating coupler. The transmitted power values are measured after the output grating coupler. The insertion loss induced by the pair of grating couplers is 21 dB in these measurements.

Table 1. Parameters used for calculating transmission spectra of the disk cavity (Device 1)

| Parameter (Symbol) | Value | Source |
|---|---|---|
| Refractive index ($n_0$, $n$) | 3.48 at 298 K<br>3.45 around 2 K | Ref. 19 |
| TPA coefficient ($\beta$) | 0.8 cm $GW^{-1}$ at 298 K<br>0.4 cm $GW^{-1}$ around 2 K | Refs. 29,30 |
| Kerr coefficient ($n_2$) | $4.5\times10^{-14}$ $cm^2$ $W^{-1}$ at 298 K<br>$2.25\times10^{-14}$ $cm^2$ $W^{-1}$ around 2 K | Refs. 29,31 |
| Thermo-optic coefficient ($\partial n/\partial T$) | $1.85\times10^{-4}$ $K^{-1}$ at 298 K<br>$< 10^{-8}$ $K^{-1}$ around 2 K | Refs. 16,19 |
| Cavity volume for TPA ($V_{TPA}$) | 0.44 $\mu m^3$ | FDTD |
| Cavity volume for free carriers ($V_{FC}$) | 0.30 $\mu m^3$ | FDTD |
| Effective mass ($m^*$) | $0.30m_0$ for electrons ($m_e^*$)<br>$0.45m_0$ for holes ($m_h^*$)<br>$m_0$: electron rest mass | Ref. 9 |
| Carrier mobility ($\mu$) | 80 $cm^2$ $V^{-1}$ $s^{-1}$ for electrons ($\mu_e$)<br>40 $cm^2$ $V^{-1}$ $s^{-1}$ for holes ($\mu_h$) | Ref. 24 |
| Carrier recombination time ($\tau_{recon}$) | 40 ns at 298 K<br>4 ns at 2.168 K<br>0.1 ns at 2.000 K | fit |
| Disk's thermal resistance ($R$) | 530 K $mW^{-1}$, in air at 298 K<br>20 K $mW^{-1}$, in LHe around 2 K | fit<br>Ref. 7 |
| Disk's resonant wavelength in linear regime ($\lambda_0$) | 1536.831 nm, in air at 298 K<br>1525.456 nm, in LHe at 2.168 K<br>1525.429 nm, in LHe at 2.000 K | measured |
| Disk's radiation quality factor ($Q_v$) | $2.33\times10^5$, in air at 298 K<br>$4.04\times10^5$, in LHe at 2.168 K<br>$4.89\times10^5$, in LHe at 2.000 K | measured |
| Disk's coupling quality factor ($Q_{in}$) | $2.10\times10^6$, in air at 298 K<br>$2.15\times10^6$, in LHe at 2.168 K<br>$2.20\times10^6$, in LHe at 2.000 K | measured |

# Supplementary Information for "Nonlinear optical effects of ultrahigh-*Q* silicon photonic nanocavities immersed in superfluid helium"

Xiankai Sun, Xufeng Zhang, Carsten Schuck, and Hong X. Tang*

*Department of Electrical Engineering, Yale University, 15 Prospect Street, New Haven, Connecticut 06511, USA*

*hong.tang@yale.edu

Here we present experimental results of a strongly overcoupled disk cavity (Device 3) characterized for representative conditions from 298 to 2 K. This device exhibits near-unity transmission (excluding the insertion loss introduced by the grating couplers) with vertical and in-plane quality factors of $Q_{\mathrm{v}} = 3.37\times10^5$ and $Q_{\mathrm{in}} = 4.93\times10^4$, respectively. Despite a lower loaded quality factor, Device 3 exhibits similar behaviors when immersed in superfluid helium as Devices 1 and 2 discussed in the main text. Accordingly, the interpretation of the nonlinear response of the disk cavity follows that provided in the main text.

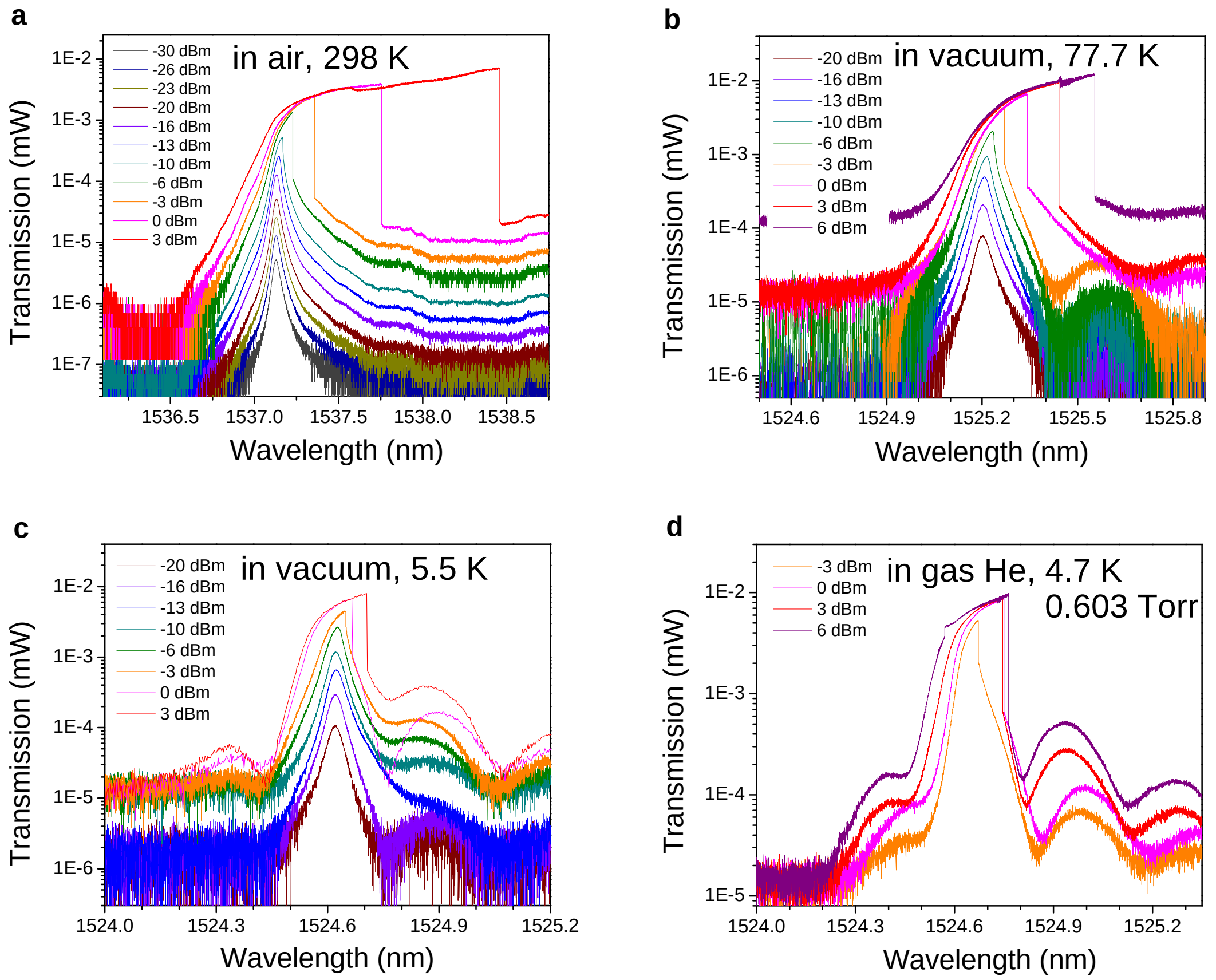

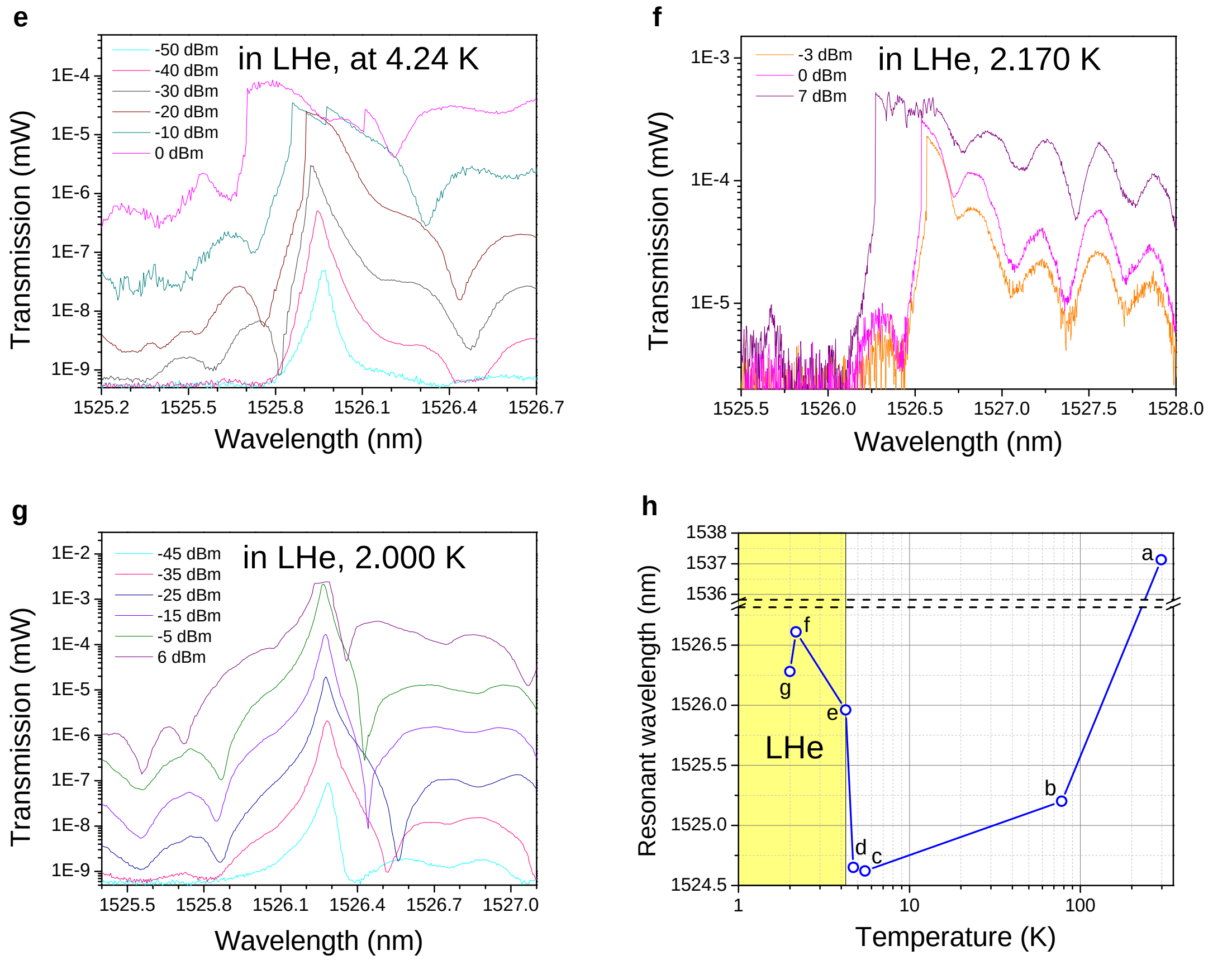

Figure S1. (a)–(g) Measured spectra of optical transmission through the disk cavity (Device 3) at various input power levels: (a) in air at 298 K (room temperature); (b) in vacuum at 77.7 K; (c) in vacuum at 5.5 K; (d) in gas helium at 4.7 K, 0.603 Torr; (e) in normal-fluid helium at 4.24 K; (f) in superfluid helium at 2.170 K; (g) in superfluid helium at 2.000 K. The quoted input power values are measured before the input grating coupler. The transmitted power values are measured after the output grating coupler. The insertion loss induced by the pair of grating couplers is 24 dB in the measurement of (a) and 21 dB in (b)–(g). (h) The disk's resonant wavelength in the linear regime under the above measurement conditions.